\documentclass[namedreferences]{solarphysics}

\usepackage[hyperref,optionalrh,showbiblabels]{spr-sola-addons} 
\usepackage{graphicx}        
\usepackage{color}           
\usepackage{breakurl}        
\usepackage[export]{adjustbox}
\usepackage{hyperref}
\hypersetup{
    colorlinks=true,
    linkcolor=red,
    filecolor=magenta,      
    urlcolor=cyan,
    citecolor=blue,
}

\newcommand{\aap}{    {\it Astron. Astrophys.}}

\newcommand{\apj}{    {\it Astrophys. J.}}
\newcommand{\apjl}{   {\it Astrophys. J. Lett.}}

\newcommand{\solphys}{{\it Solar Phys.}}
 
\newcommand{\ssr}{    {\it Space Sci. Rev.}} 
\newcommand{\pasa}{    {\it Publications of the Astronomical Society of Australia}} 

\chardef\us=`\_

\begin{document}

\begin{article}
\begin{opening}

\title{Type IV Radio Bursts and Associated Active Regions in the Sunspot Cycle 24\\ {\it Solar Physics}}

\author[addressref={aff1},corref,email={anshu.kumari@helsinki.fi}]{\inits{Anshu}\fnm{Anshu Kumari \orcid{0000-0001-5742-9033}}~\lnm{}}

\address[id=aff1]{Department of Physics, University of Helsinki, P.O. Box 64, FI-00014 Helsinki, Finland}

\runningauthor{Anshu Kumari}
\runningtitle{Type IV bursts and Active regions}

\begin{abstract}
   
In this article, the association of solar radio type IV bursts with active region location on the Sun is studied for the solar cycle 24. 
The active regions associated with moving and stationary type IV bursts are categorised as close to disk center and far from disk center regions based on their location on the solar surface (i.e, $\leq 45^{\circ}$ or $\geq 45^{\circ}$, respectively). 
The location of the active regions associated with type IV bursts accompanied with coronal mass ejections (CMEs) are also studied. 
We found that $\approx 30-40 \%$ of the active regions are located far from disk center for all the bursts. 
It is found that most of the active regions associated with stationary type IV bursts are close to disk center ($\approx 60-70 \%$). The active regions associated with moving type IV bursts are more evenly distributed across the surface, i.e $\approx 56 \%$ and $\approx 44 \%$, close to disk center and far from disk center regions, respectively.
Most of the burst having active region close to disk center indicate that these bursts can be used to  obtain physical properties such as electron density and magnetic fields of the coronal mass ejections responsible for geomagnetic storms.

\end{abstract}
\keywords{Corona, radio bursts, type IV, coronal mass ejections, active region}
\end{opening}

\section{Introduction}
     \label{sec:section1} 

The Sun, being our nearest star, has significant effects on the near-Earth atmosphere, i.e., space weather. The large-scale explosion on the Sun, known as coronal mass ejections (CMEs) \citep[][]{howard1985coronal, Yashiro2004, Cho2007, Vourlidas2010, Webb2012} are often seen in white-light coronagraph images taken with space and ground-based instruments \citep[][]{fisher1981new, Brueckner1995, Howard2008}. The solar radio type II and type IV bursts are of particular interest for space weather as their studies can help to understand the near-Sun development of interplanetary disturbances \citep[][]{Gary1985,pohjolainen2008cme}.

Type II bursts originate at the shock in front of a CMEs and are plasma emission \citep[][]{Smerd1975, Mann1995, Aurass1997, gopalswamy2005type, Anshu2017a,  kumari2017b}. The type II radio bursts, their near Sun signatures and their source regions have been very well studied previously \citep[e.g.,][]{Roberts1959, cliver1999origin, Vrsnak2002, Gopalswamy2006, Ramesh2010a, kumari2017c,  kumari2019direct}. 
Type IV radio bursts are most frequently associated with CME flux ropes or the footpoints of the magnetic loops \citep[][]{Riddle1970, Ramesh2004, Vasanth2019, salas2020polarisation}. These bursts can have variable emission mechanisms, unlike type II bursts, eg. plasma emission \citep[][]{Gary1985, Hariharan2016a,  Liu2018, Morosan2019} or gryo-synchrotron emission \citep[][]{Bain2014, Sasikumar2014, Carley2017}. 
Type II and type IV bursts can be used to estimate the coronal magnetic fields ($B$) associated with various parts of the CME. For example, \cite{Gopalswamy2012} estimated the $B$ fields values as $\approx$ 1.5--1.1 $G$ at $\approx$ 1.3-1.5 $R_{\odot}$ using shock standoff method and splitband type II bursts, whereas \cite{Sasikumar2014} estimated the $B$ values as $\approx $ 1.4--2.2 $G$ at $\approx$ 1.9-2.2 $1.4 R_{\odot}$. Several other authors reported similar values in the same height range
\citep[see e.g.,][and the references therein]{Cho2007, Tun2013, Sasikumar2014, kumari2017b, kumari2019direct}.

Long-term statistical studies of type IV bursts and their association with CMEs/active regions (ARs) are less explored in comparison to type II bursts \citep[][]{lara2003statistical, Kahler2019}. \cite{Robinson1978} and \cite{Gergely1986} studied few type IV bursts and their relation with CMEs. Recently, \cite{Kumari_2021} studied the association on these bursts with CMEs in the previous solar cycle (cycle 24) based on the spectral properties of these bursts. \cite{Morosan_2021} studied the moving bursts in the rising phase of solar cycle 24 with the usefulness of using radio images.
However, due to the lack of radio imaging instruments, locating the source region of these radio bursts in the corona becomes almost impossible. One way to locate the initial source of any radio bursts is to identify the associated active region. Hence, in this work we study the association of active regions with type IV radio bursts. The location of active regions associated with type IV bursts are investigated, and their association with CMEs is studied. Depending on their spectral features, type IV radio bursts are also classified as moving and stationary bursts.
The paper is arranged as follows: in section \ref{sec:section2}, the observational data used for the study is described; in section \ref{sec:section3} the data analysis method is explained; and in section \ref{sec:section4} the results obtained in the study is discussed and the paper is concluded. 
 
 
\section{Observational Data}
     \label{sec:section2} 

\begin{figure}
\centering\includegraphics[width=1\textwidth,clip=]{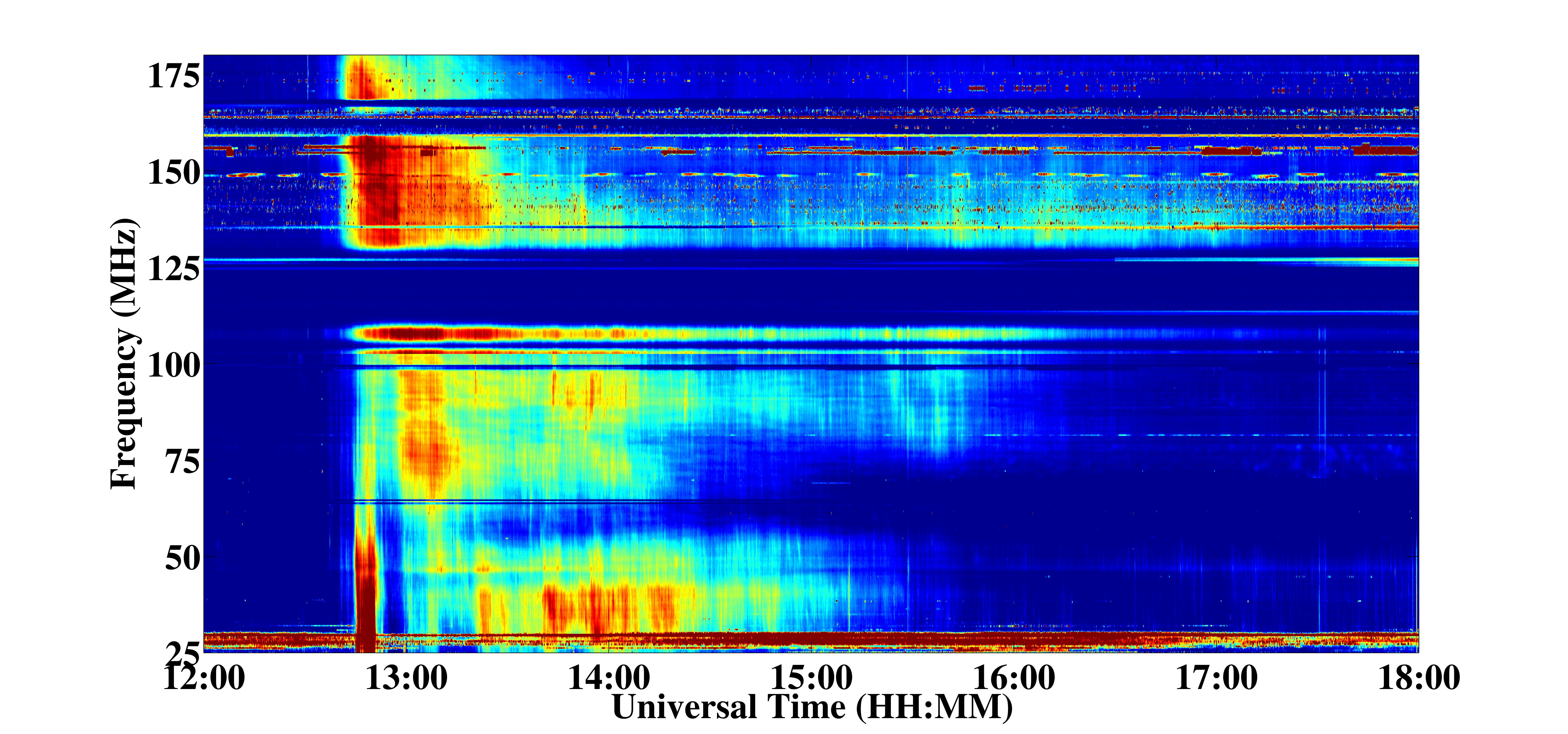}
\caption{Solar radio dynamic spectra of a type IV burst recorded at the Sagamore Hill Solar Radio Observatory on September 24, 2011. This type IV burst started at $\approx$ 12:50 UT and lasted till $\approx$ 17:40 UT. The frequency range of this burst was $\approx$ 180-25 MHz. The horizontal stripes in the dynamic spectra are due to the local radio frequency interference (RFIs).}
\label{fig:figure1}
\end{figure}

The event lists available at the Space Weather Prediction Center (SWPC)\footnote{\url{https://www.swpc.noaa.gov/products/solar-and-geophysical-event-reports}} was extensively used for this study. SWPC provides the list of all the solar and geophysical Event Reports every day since 1996\footnote{\url{ftp://ftp.swpc.noaa.gov/pub/warehouse/}}. 
The list contains X-ray events, optical flares, radio bursts, and their intensity, type, impact, associated active regions (ARs), etc. Several observatories report the events, and it is compiled, combined, and updated on SWPC website. For radio events, the Radio Solar Telescope Network (RSTN) and a couple of other observatories contribute in reporting the bursts. The radio bursts are classified based on their spectral and temporal appearance in the solar dynamic spectra. The type IV bursts information from January 01, 2009 - December, 31, 2019 were extracted from the entire event list. It contains: the start and end time, start and end frequency, reporting station and associated AR for the type IV bursts. Figure \ref{fig:figure1} shows dynamic spectra of type IV burst on September 24, 2011, with the Sagamore Hill Solar Radio Observatory.
The start time of this burst was $\approx$ 12:50 UT, and it lasted for $\approx$ 4 hours. The frequency range of this stationary type IV bursts was $\approx 25-180 $ MHz.

For the CME association, we used the list provided at the coordinated data analysis website \citep[CDAW;][]{Yashiro2004}\footnote{\url{https://cdaw.gsfc.nasa.gov/CME_list/index.html}}, which contains the manual detection of CMEs with the Large Angle and Spectrometric Coronagraph (LASCO) onboard the Solar and Heliospheric Observatory \citep[SOHO;][]{Brueckner1995}.
We also used CMEs detected with the Cor1 and Cor2 coronagraphs from the Sun-Earth Connection Coronal and Heliospheric Investigation \citep[SECCHI;][]{Howard2008} onboard the Solar Terrestrial Relations Observatory (STEREO).
We used the list provided on the Solar Eruptive Event Detection System (SEEDS)\footnote{\url{http://spaceweather.gmu.edu/seeds/secchi/detection_cor2/monthly/}} website and the CME list provided by \citet{Vourlidas2017}, where the authors used the  dual-viewpoint CME catalog from the SECCHI/COR telescopes\footnote{\url{http://solar.jhuapl.edu/Data-Products/COR-CME-Catalog.php}}.

\section{Methodology and Implementation}
     \label{sec:section3} 

The aim of the work is to study the active region source locations of the radio type IV bursts during the solar cycle 24. The SWPC daily event list was used to obtain the active region number and location associated with the type IV bursts. This list assigns the active region from optical images/GOES Solar X-ray Imager to the radio bursts. Those bursts which had AR source $\leq 45^{\circ}$ were classified as `close to disk center' region bursts, and the remaining were classified as  `far from disk center' region bursts. The classification of type IV radio bursts as moving and stationary (type IVm and type IVs, respectively) was based on the burst appearance in the dynamic spectra \citep[for detail, see][]{Kumari_2021}. 
We used the drift rate and duration of the type IV bursts to classify the bursts as type IVm and type IVs. A constant drift rate was calculated for both stationary and moving type IV bursts by using the start and end frequency of the bursts, and the duration of the bursts as, $DR = \frac{F_H-F_L}{t}$, where $DR$, $F_H, F_L$ and $t$ are the drift rate, start frequency, end frequency and duration of the bursts, respectively. The bursts with $DR \geq 0.03 $ MHz/s were classified as moving bursts \citep[][]{Robinson1978, Gergely1986, Kumari_2021}. 
The bursts with lower drift rates ($DR < 0.03 $ MHz/s) were classified as stationary bursts.
\cite{Robinson1978} and \cite{Gergely1986} had earlier reported that the moving type IV bursts have duration less than an hour. We used this criteria to further distinguish between stationary and moving type IV bursts for the bursts which were ambiguous to classify only using DRs.
Figure \ref{fig:figure1} shows dynamic spectra of a stationary type IV burst observed on September 24, 2011.


The CME association of these radio bursts was checked with the CMEs detected with SOHO/LASCO-C2, STEREO-A and B, on a temporal correlation \citep[for detail, see][]{Kumari_2021}. 
The first appearance of the CME in coronagraph field of view (FOV) depends upon the speed, acceleration, and the propagation direction of the CME with respect to the Sun-Earth line (for LASCO) and the Sun-STEREO line (for STEREO). For the CME-type IV association, we had two criteria as, i) the active region associated with the CME and type IV should be the same (this was done by comparing the position angle (PA) of the CME with AR related to the type IV burst); ii) a CME should appear in LASCO-C2 FOV within  $\approx 2 $ hours of the start time of the type IV burst (we note that time between CME and type IV association can vary depending on the location of the ejection, for example, a limb CME may appear early in LASCO-C2 FOV). Both of these criteria had to be fulfilled to assure that the right pair of white-light CME and radio signature were connected.
Table \ref{tab:table1} contains the number and percentage of type IV bursts with their AR locations. It also shows the number and percentage of type IV bursts after classification as moving and stationary. 
The table also lists the type IV bursts and their association with identified CMEs with SOHO/LASCO and STEREO-A and B coronagraphs.

\section{Results and Discussion}
     \label{sec:section4}

\begin{figure}
\centering\includegraphics[width=0.32\textwidth, trim={3cm 0 3cm 0}, clip=]{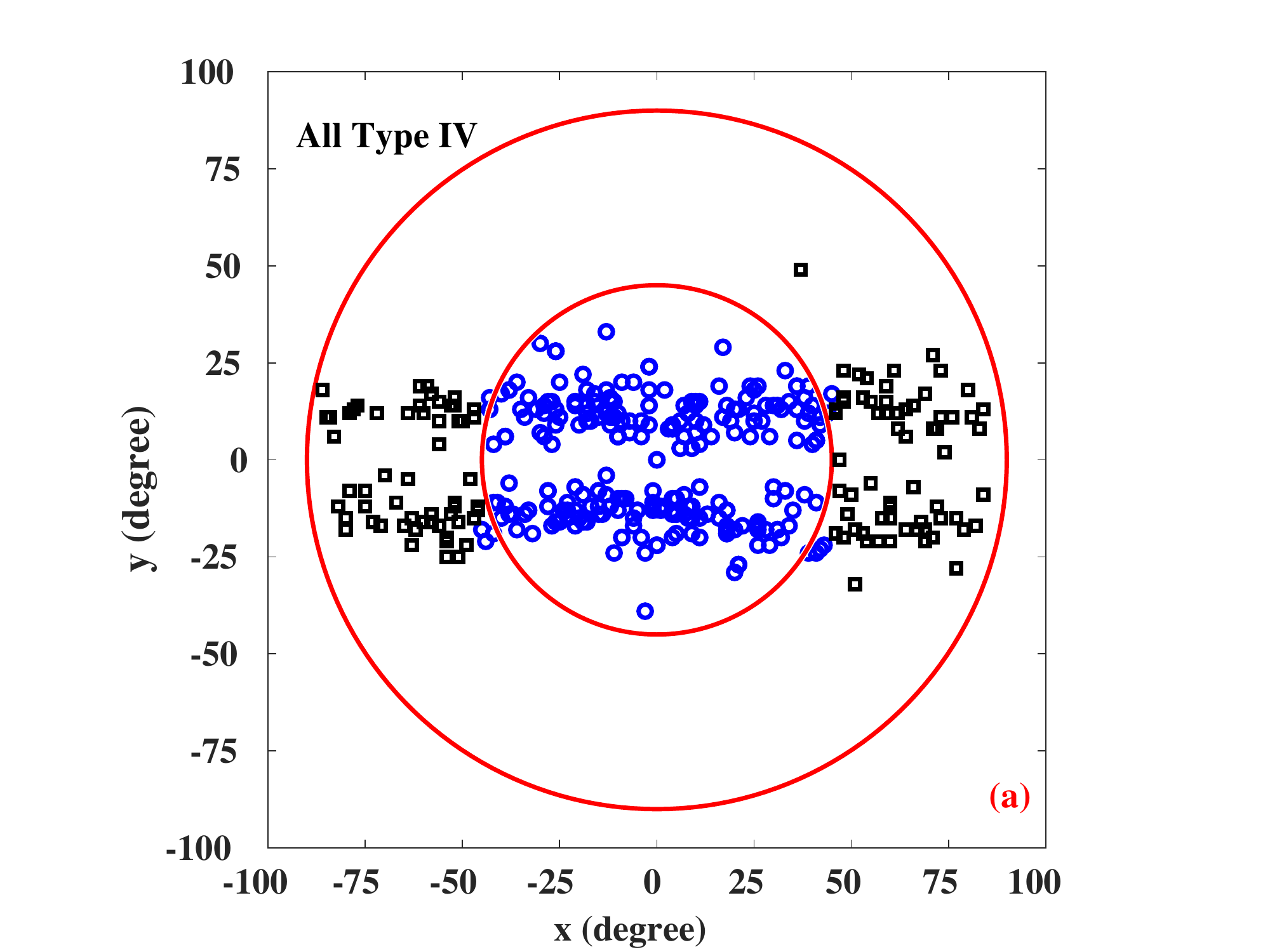}
\centering\includegraphics[width=0.32\textwidth, trim={3cm 0 3cm 0}, clip=]{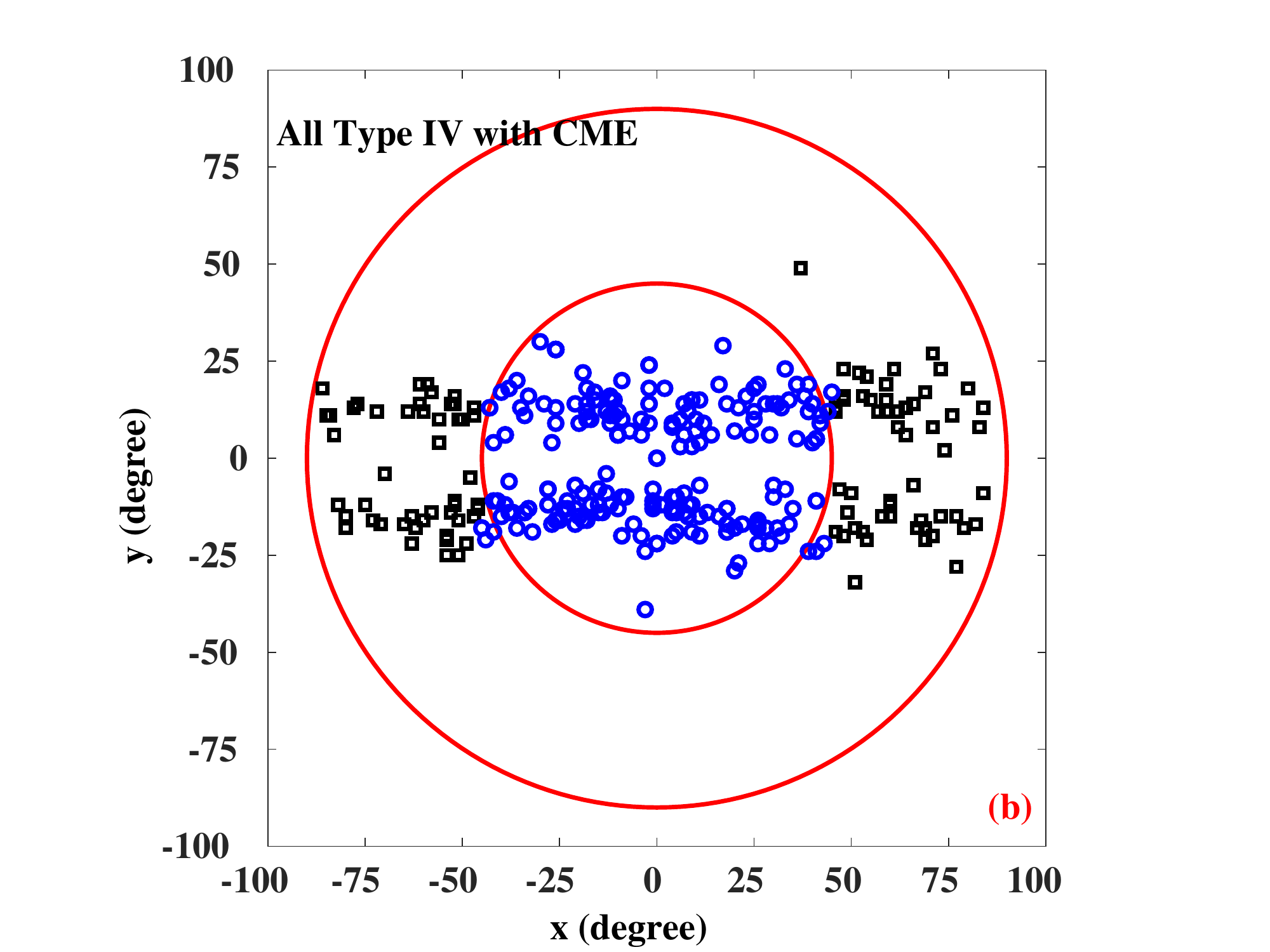}
\centering\includegraphics[width=0.32\textwidth, trim={3cm 0 3cm 0},clip=]{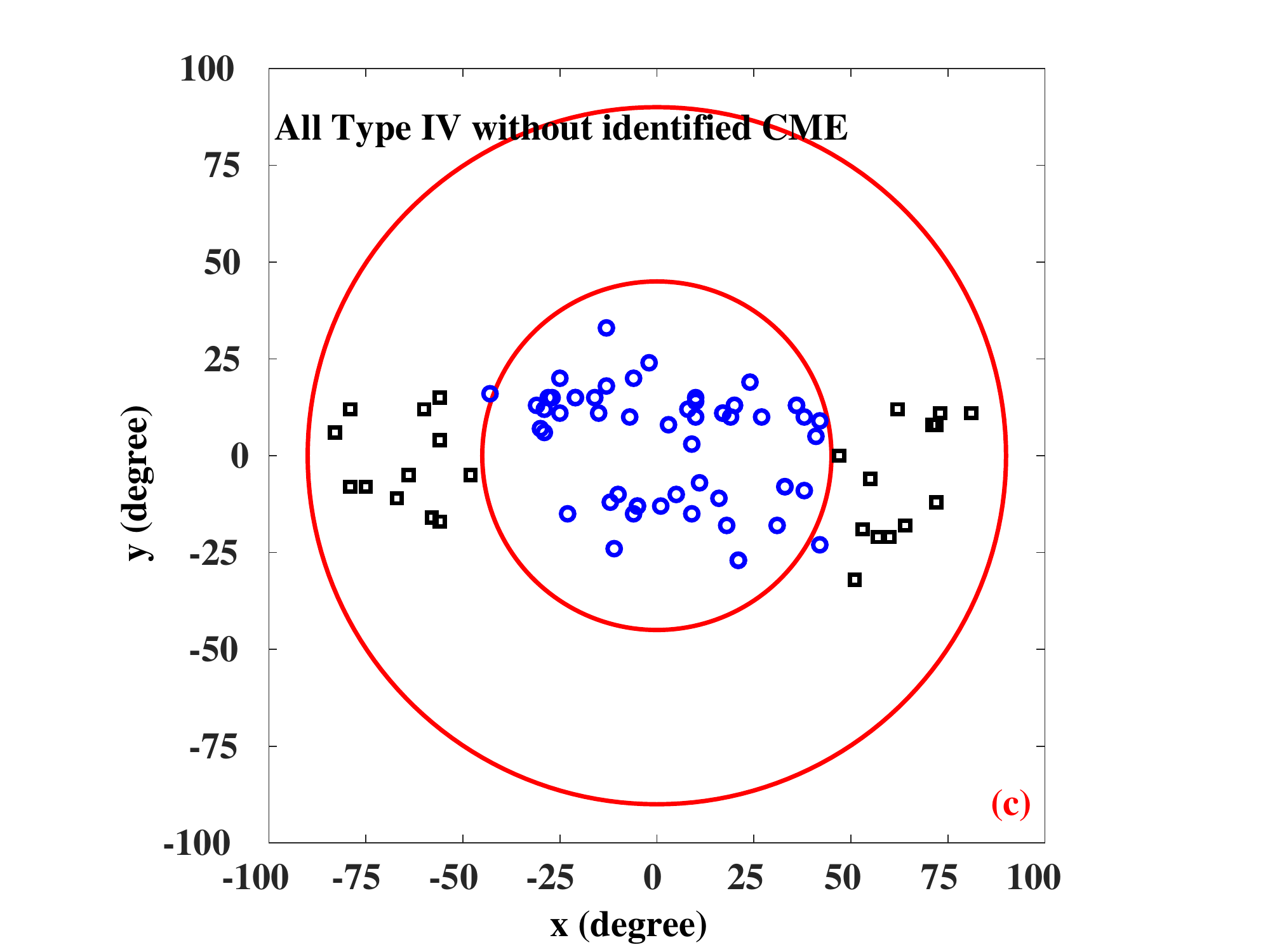}
\centering\includegraphics[width=0.32\textwidth,trim={3cm 0 3cm 0}, clip=]{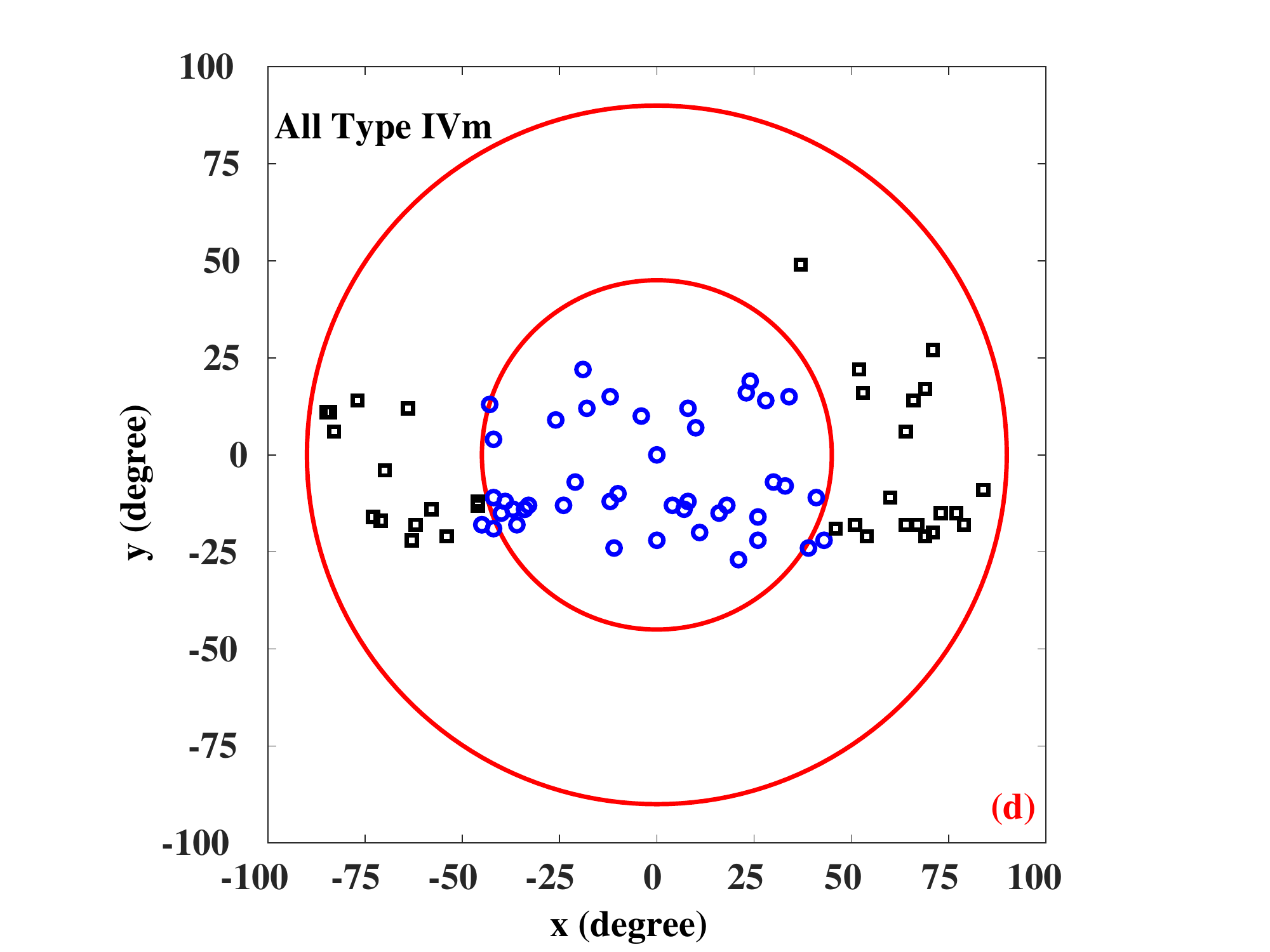}
\centering\includegraphics[width=0.32\textwidth,trim={3cm 0 3cm 0}, clip=]{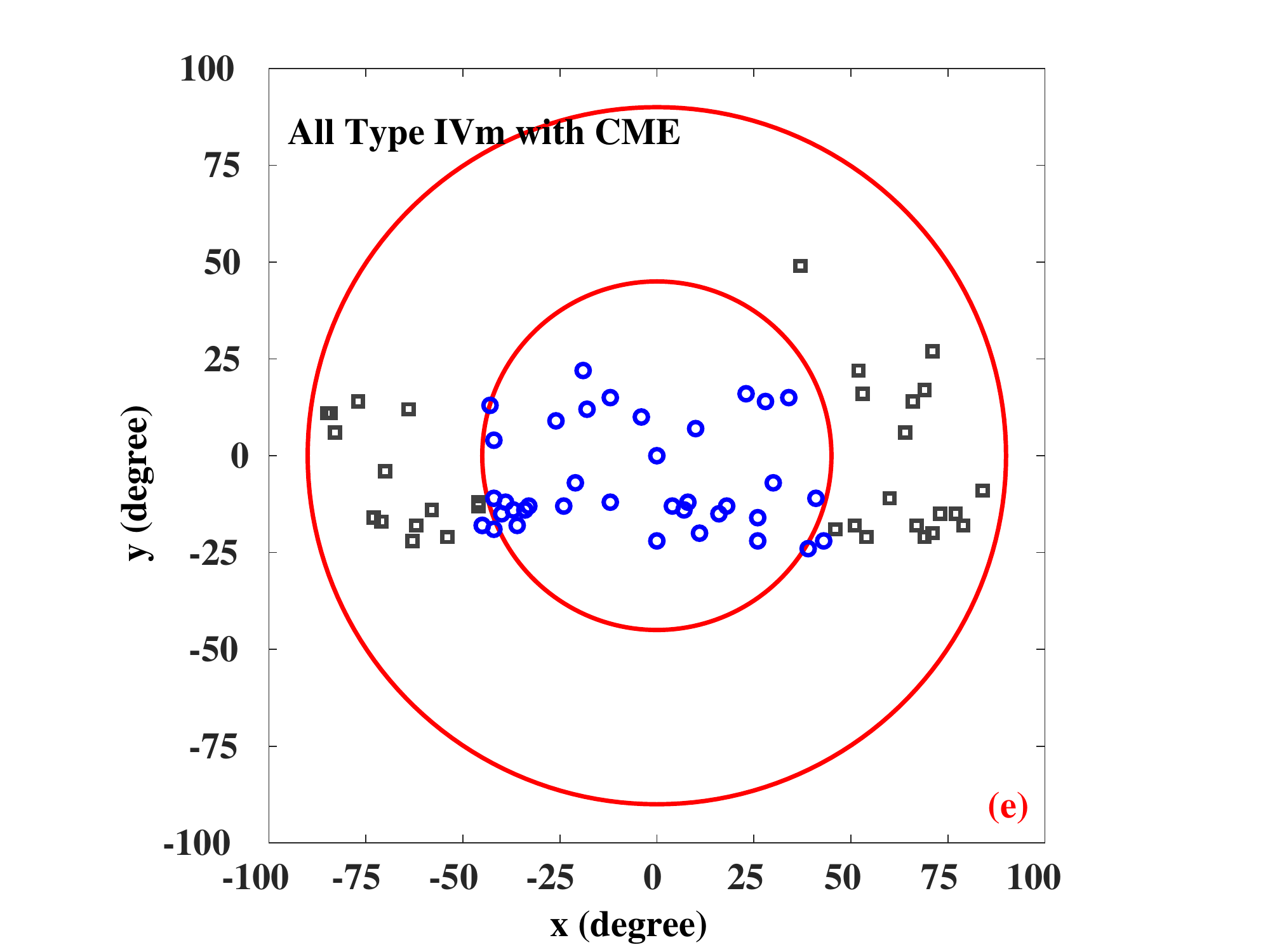}
\centering\includegraphics[width=0.32\textwidth,trim={3cm 0 3cm 0}, clip=]{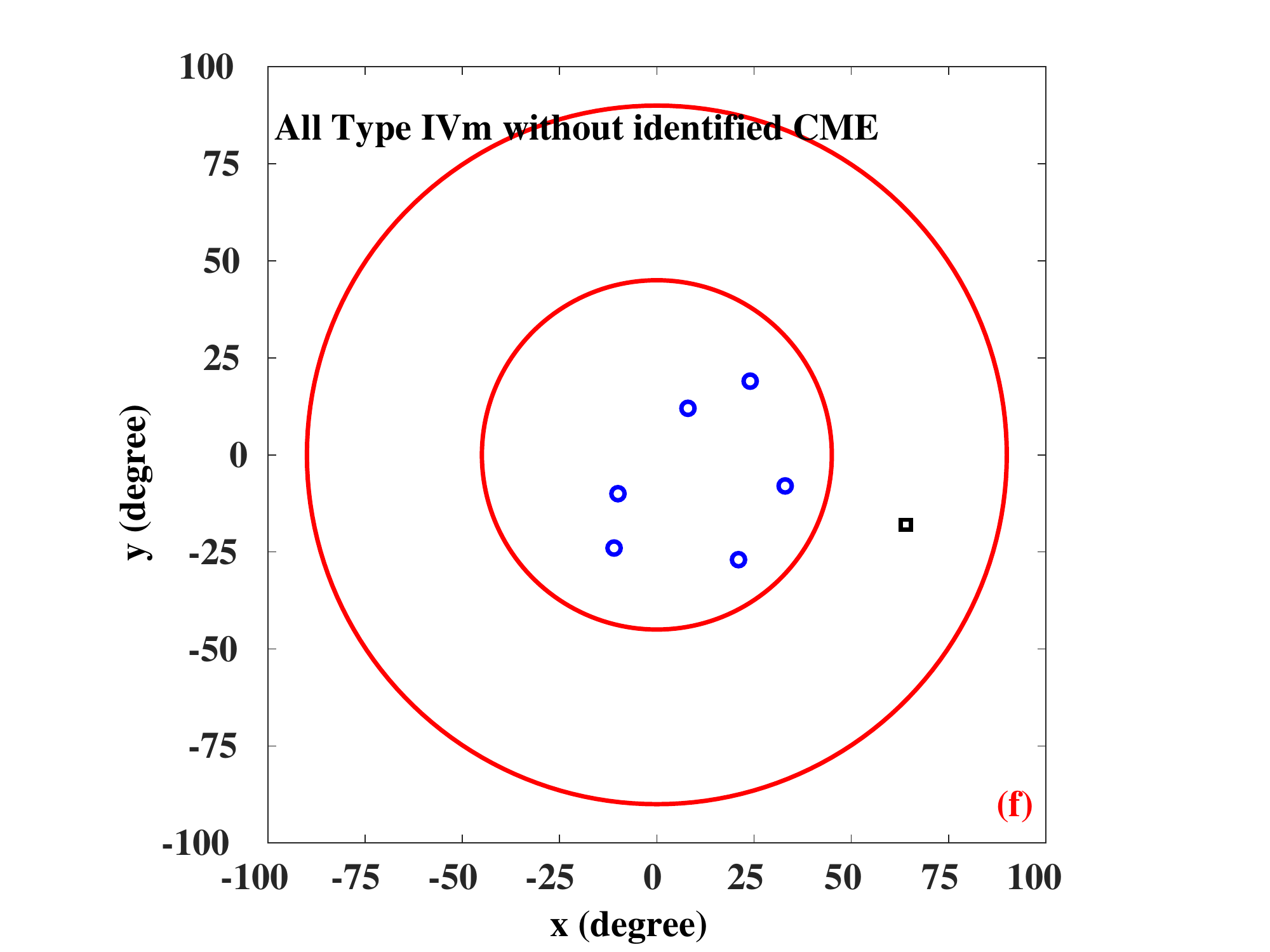}
\centering\includegraphics[width=0.32\textwidth,trim={3cm 0 3cm 0}, clip=]{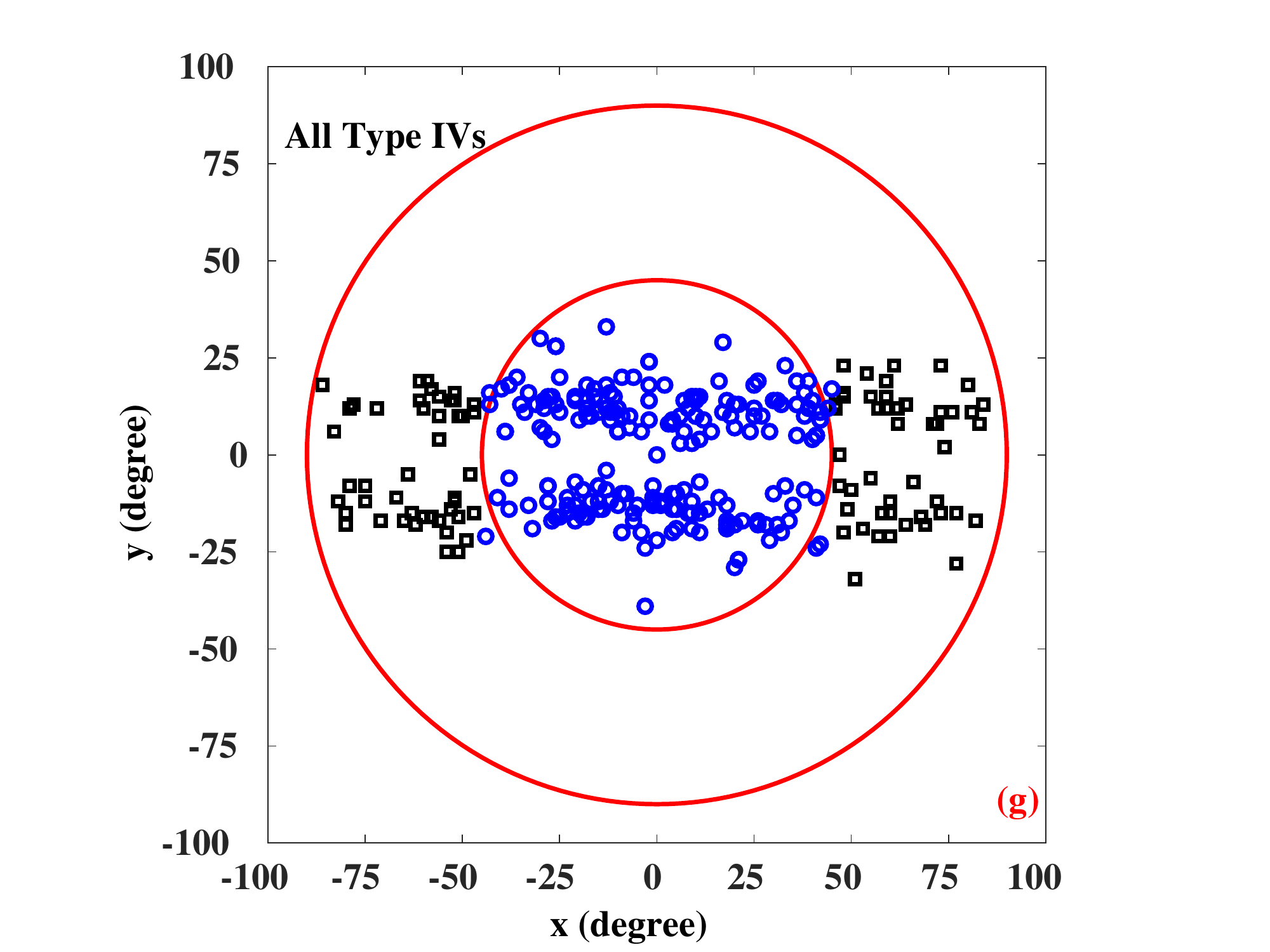}
\centering\includegraphics[width=0.32\textwidth,trim={3cm 0 3cm 0}, clip=]{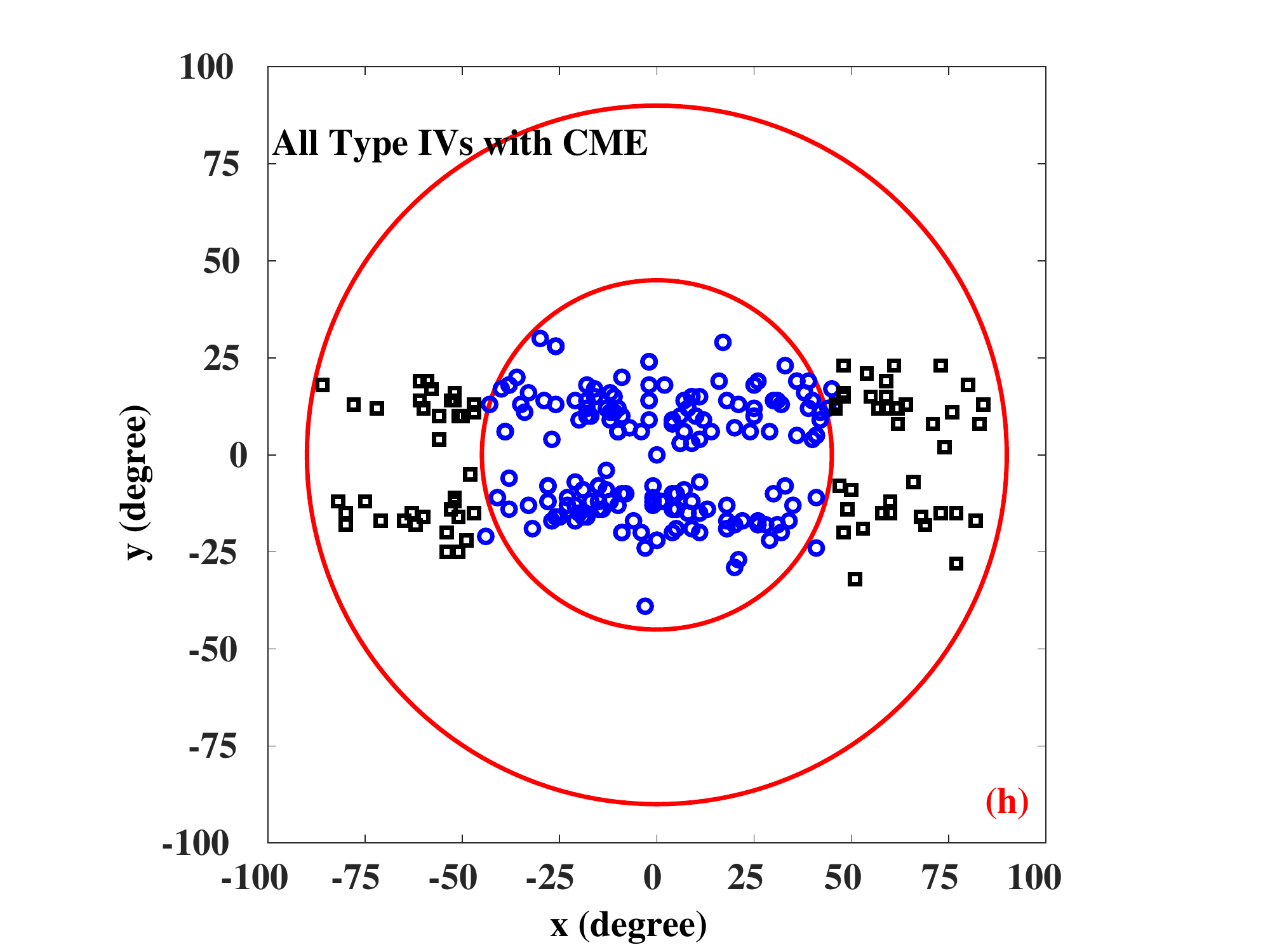}
\centering\includegraphics[width=0.32\textwidth,trim={3cm 0 3cm 0}, clip=]{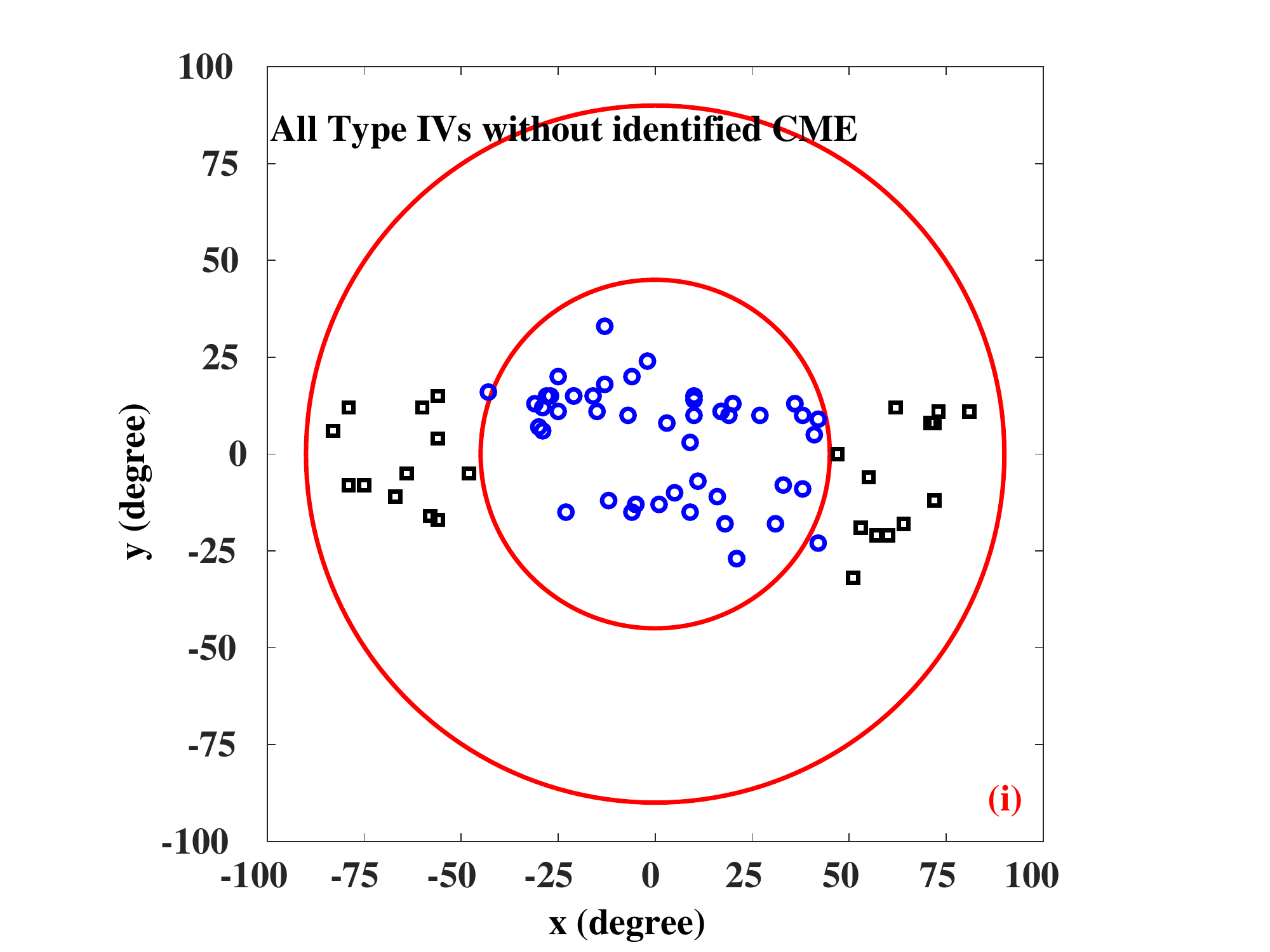}
\caption{The active region location of the: 
(a) All type IV bursts;
(b) All type IV bursts with CMEs;
(c) All type IV bursts without identified CMEs;
(d) All moving type IV bursts;
(e) All moving type IV bursts with identified CMEs;
(f) All moving type IV bursts without identified CMEs;
(g) All stationary type IV bursts;
(h) All stationary  type IV bursts with CMEs;
(i) All stationary  type IV bursts without identified CMEs.
The `blue' and `black' points represents close to and far from the disk center locations, respectively. The `inner' and `outer' red circles represent the $45^{\circ}$ mark and the solar photosphere, respectively. The axes are marked in coordinates from $-90^{\circ}$ to $90^{\circ}$ from South to North and East to West on the solar surface.}
\label{fig:figure2}
\end{figure}

Figure \ref{fig:figure2} shows the location of active regions with which the type IV radio bursts were associated. 
Out of the 446 type IV bursts in solar cycle 24, 291 bursts ($\approx 65 \%$) were having AR $\leq 45^{\circ}$ (see Table \ref{tab:table1} for details), hence they were called bursts associated with close to disk center AR.
The rest of the bursts were associated with AR far from disk center ($\approx 35 \%$ ).
For the type IV burst accompanied with CMEs, almost same number of burst ($\approx 65 \%$) were located close to disk center. Only $\approx 35 \%$ of the bursts had AR far from disk center. The bursts which were not associated with CMEs had flare association \citep[][]{Kumari_2021}. 
\cite{Gopalswamy2011b} suggested that sometimes the type IV emission can have a connection to the flare reconnection site. Type IV emissions can also come from from either one or both the legs of the magnetic flux ropes, which could also be associated with a failed eruption. \cite{labrum1985solar} had summarised that non-thermal electrons related to flares can be there in the corona for several hours, hence we can observe the stationary type IV bursts with any non identified CME. 
We note that between 2014 to 2016, the STEREO-A and B were close to the Sun-Earth line. As faint CMEs traveling close to the observer's line of sight may not be observed in coronagraphs, hence these coronagraphs and LASCO-C2 would have failed to detect CMEs arising from close to disk center.
This could increase the number of type IV bursts without identified CMEs in our list, specially during the above mentioned period. For the moving bursts, the AR location were almost equally distributed between close to disk center ($\approx 56 \%$) and far from disk center ($\approx 44 \%$). Only one type IVm burst without any identified CME has originated far from the disk center in cycle 24. 
The AR location distribution of stationary type IV bursts showed almost identical distribution like all the type IV bursts, with $\approx 67 \%$ and $\approx 33 \%$, close to and far from disk center, respectively.

\begin{table}
\caption{Type IV bursts and the  associated active regions }
\label{tab:table1}
\begin{tabular}{|lccccc|}
\hline
Description          & All & Close to & \% Close to & Far from & \% Far from \\ 
                     & All & Disk center & Disk center & Disk center & Disk center \\ 
\hline
Type IV              & 446 & 291  & 65 \%   & 155  & 35 \%   \\
Type IV with CME     & 359 & 234  & 65 \%   & 125  & 35 \%   \\
Type IV without identified CME  & 87  & 57   & 66 \%   & 30   & 34 \%   \\
\hline
Type IVm             & 80  & 45   & 56 \%   & 35   & 44 \%   \\
Type IVm with CME    & 73  & 39   & 53 \%   & 34   & 47 \%   \\
Type IVm without identified CME & 7   & 6    & 86 \%   & 1    & 14 \%   \\
\hline
Type IVs             & 366 & 246  & 67 \%   & 120  & 33 \%   \\
Type IVs with CME    & 286 & 195  & 68 \%   & 91   & 32 \%   \\
Type IVs without identified CME & 80  & 51   & 64 \%   & 29   & 36 \%  \\
\hline
\end{tabular}
\end{table}

\begin{figure}
\centering \includegraphics[width=0.8\textwidth,clip=,]{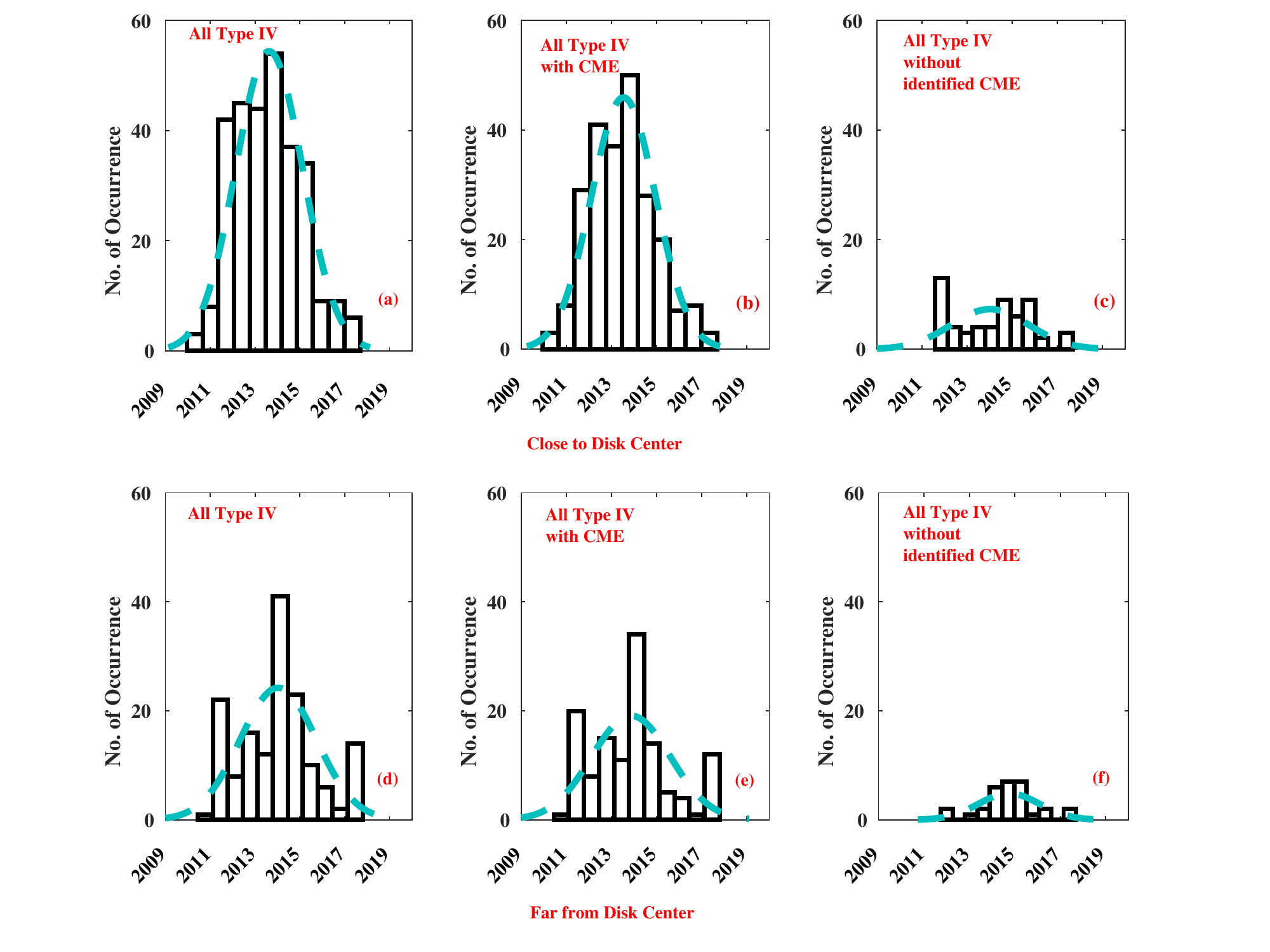}
\centering \includegraphics[width=0.8\textwidth,clip=, ]{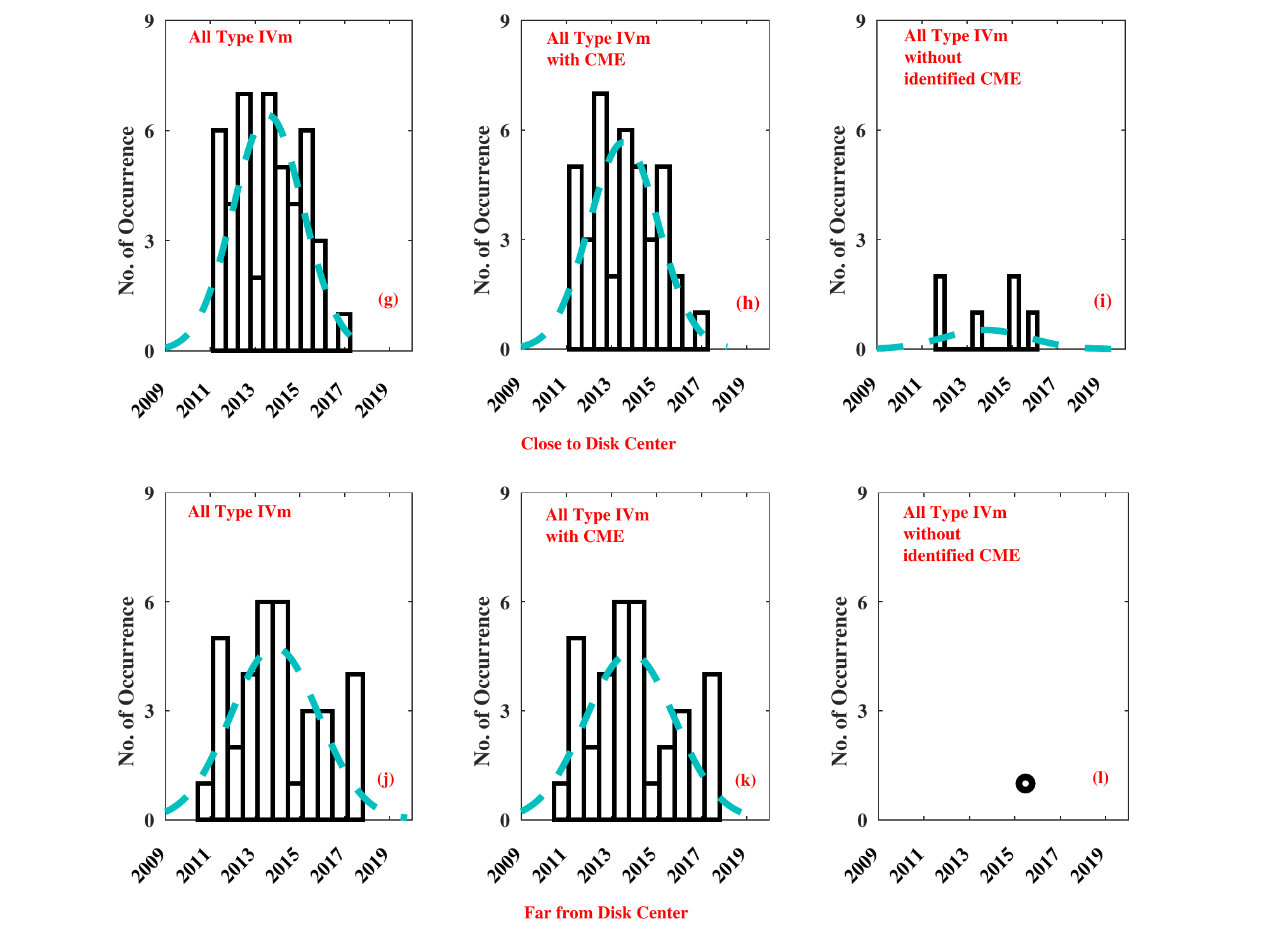}
\centering \includegraphics[width=0.8\textwidth,clip=, ]{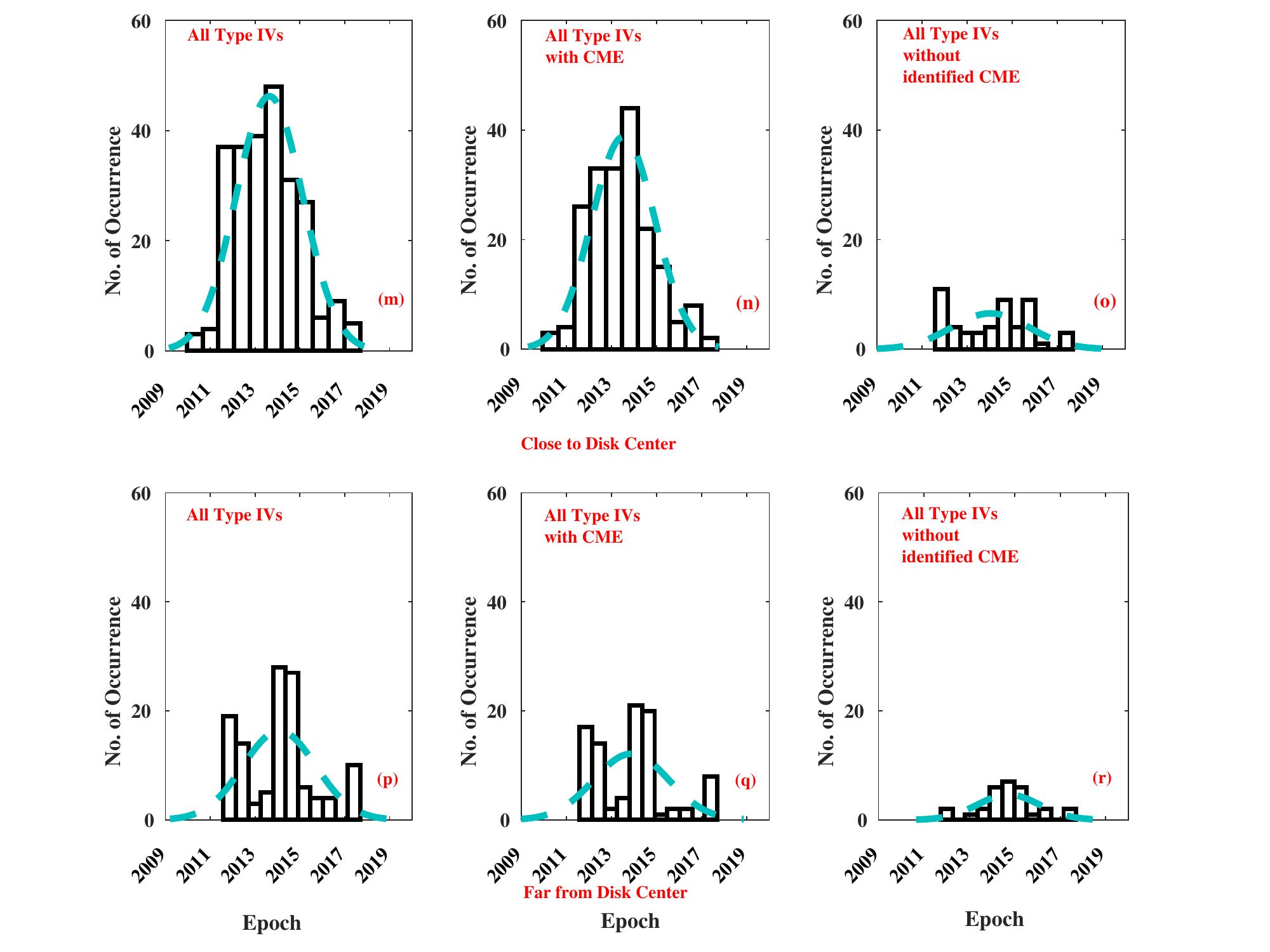}
\caption{Histograms showing the variation of the active region location associated with different type of type IV bursts in the solar cycle 24.
The first two rows of plots correspond to all type IV events, the middle two rows to moving type IV bursts and the last two rows to stationary type IV bursts.}
The cyan dashed lines are Gaussian fits to the histograms.
\label{fig:figure3}
\end{figure}

Figure \ref{fig:figure3} shows the year-wise distribution of the associated active region with the type IV bursts. This figure also shows the histograms of moving and stationary type IV bursts close to and far from the disk center, along with their CME association.
We used Gaussian fits for the histograms. For all the Gaussian fits, the peaks lie between 2013-2014, which was also the peak for solar cycle 24. This indicates that the occurrence of the type IV bursts was directly dependent on the solar cycle variation. The fits indicate that $\approx 20-40 \%$ of bursts occurred during the solar maxima. The only moving type IV burst which had AR far from the disk center occurred during the declining phase of cycle 24. We noticed that there there were more type IV bursts during the rise period of the solar cycle 24 than during the decline phase of the cycle. 

There were no type IVm bursts during the beginning years of cycle 24 (see Figure \ref{fig:figure3}g), which were associated with AR close to disk center. Similar to this, no type IVs bursts originated from AR far from disk center during the start of the solar cycle 24. 
Almost all the moving type IV bursts without any identified CME association had originated from the AR close to the disk (refer Figure \ref{fig:figure2}f). 
During the end of cycle 24, in the year 2018-2019, there were no type IV bursts observed.

Metric-type II and deca-hectometer (DH)-type II radio bursts are often used as an early indicator of CMEs and interplanetary CMEs (ICMEs). These DH-type II bursts indicate the MHD shock entering the interplanetary medium. The CMEs associated with the active region near to the central meridian (the disk) are often responsible for geomagnetic storms, as those are in the line of sight of the Earth \citep[see e.g.,][and references therein]{Schwenn2005, Gopalswamy2010, Cid2014}. Since the type IV bursts are also associated with CMEs, these bursts can help to understand the near-Sun development of interplanetary disturbances \citep[][]{Gary1985}.
Of lately, authors have also used type IV radio bursts to extract near-Sun CME parameters. \cite{7601385} showed the moving type IV bursts are associated with the moving flux ropes of the CMEs, hence can be used as a proxy to determine the electron density and magnetic fields of these CMEs \citep[][]{Gergely1986}. 
Since the magnetic field is responsible for the CME eruption, evolution, and later the geo-effectiveness, moving type IV bursts can aid in the determination of the latter \citep[e.g.,][]{Bastian2001,Tun2013, Carley2017}. 
It has been previously reported by \cite{gonzalez1990dual} and \cite{Echer2011} that the first peak in the two-peak variation of geomagnetic activities in previous solar cycles was due to CMEs. 
In this study, we found that that moving type IV bursts are almost exclusively associated with CMEs ($\approx 90 \%$), they can be widely used for estimating the geo-effective CME magnetic fields.
\cite{Kumari_2021} reported high association of type IV bursts with CMEs. It indicates if a type IV burst is present during a CME, it can be used for studying the electron acceleration locations as well as CME kinematics early during the eruption process. 
\cite{salas2020polarisation} studied stationary type IV radio bursts, where they showed that the stationary type IV bursts originate from the legs of the magnetic flux rope erupting into the high corona during the CME. This would indicate that a CME eruption is necessary for the generation of stationary type IV emission. A high association of stationary type IV bursts with CMEs indicates that type IV bursts can be used to estimate the magnetic field near the legs of the erupting flux ropes.

\cite{Zhang2003} found that the most geo-effective CMEs originate from the western hemisphere. Geo-effective CMEs mainly originate within E20 to W50 longitude, i.e., close to the disk center \citep[][]{Zhang2003}. We found that most of the type IV bursts were associated with AR close to the disk center ($\approx 65 \%$), hence can be used to determine the physical parameters of at least some of the geo-effective CMEs near the Sun. 
With powerful radio instruments like the upgraded Nan{\c c}ay Radioheliograph \citep[NRH;][]{Kerdraon1997}, the LOw Frequency ARray \citep[LOFAR;][]{Haarlem2013}, the Murchison Widefield Array \citep[MWA;][]{Tingay2013}, and the Gauribidanur RAdioheliograPH \citep[GRAPH;][]{ramesh1998} this type of study can further be benefited where the radio source region of radio bursts can be identified using radio imaging. 

\begin{acks}
The SOHO/LASCO data used here are produced by a consortium of the Naval Research Laboratory (USA), Max-Planck-Institut fuer Aeronomie (Germany), Laboratoire d'Astronomie (France), and the University of Birmingham (UK). SOHO is a project of international cooperation between ESA and NASA. The SECCHI data used here were produced by an international consortium of the Naval Research Laboratory (USA), Lockheed Martin Solar and Astrophysics Lab (USA), NASA Goddard Space Flight Center (USA), Rutherford Appleton Laboratory (UK), University of Birmingham (UK), Max-Planck-Institut for Solar System Research (Germany), Centre Spatiale de Liège (Belgium), Institut d’Optique Théorique et Appliquée (France), Institut d’Astrophysique Spatiale (France). 
A.K acknowledges Prof. R Ramesh, Dr. C. Kathiravan, Prof. Emilia Kiplua and Dr. Diana Morosan for fruitful discussions. A.K. also acknowledges the anonymous referee for the useful suggestions and comments.
This research made use of NASA’s Astrophysics Data System (ADS). 
A.K. acknowledges the ERC under the European Union's Horizon 2020 Research and Innovation Programme Project SolMAG 724391.  
\end{acks}
\vspace{1cm}

{\textbf{Data Availability:} The datasets generated during and/or analysed during the current study are available from the corresponding author on reasonable request.}

\bibliographystyle{spr-mp-sola}


\end{article} 

\end{document}